\documentclass[fleqn,10pt]{wlscirep}
\usepackage[utf8]{inputenc}
\usepackage[T1]{fontenc}
\usepackage{wrapfig}
\usepackage{verbatim}
\usepackage{siunitx}
\usepackage{hyperref}
\usepackage{subcaption}
\title{Motif Caller: Sequence Reconstruction for motif-based DNA Storage}

\author[1, *]{Parv Agarwal}
\author[2]{Nimesh Pinnamaneni}
\author[1]{Thomas Heinis}

\affil[1]{Department of Computing, Imperial College London, United Kingdom}
\affil[2]{Helixworks Technologies, Cork, Ireland}
\affil[*]{p.agarwal23@imperial.ac.uk}
\begin{abstract}
DNA data storage is rapidly emerging as a promising solution for long-term data archiving, largely due to its exceptional durability. However, the synthesis of DNA strands remains a significant bottleneck in terms of cost and speed. To address this, new methods have been developed that encode information by concatenating long data-carrying DNA sequences from pre-synthesized DNA subsequences — known as motifs — from a library.

Reading back data from DNA storage relies on basecalling—the process of translating raw nanopore sequencing signals into DNA base sequences using machine learning models. These sequences are then decoded back into binary data. However, current basecalling approaches are not optimized for decoding motif-carrying DNA: they first predict individual bases from the raw signal and only afterward attempt to identify higher-level motifs. This two-step, motif-agnostic process is both imprecise and inefficient.

In this paper we introduce \emph{Motif Caller},  machine learning model designed to directly detect entire motifs from raw nanopore signals, bypassing the need for intermediate basecalling. By targeting motifs directly, \emph{Motif Caller} leverages richer signal features associated with each motif, resulting in significantly improved accuracy. This direct approach also enhances the efficiency of data retrieval in motif-based DNA storage systems.
\end{abstract}
\begin{document}

\flushbottom
\maketitle
\thispagestyle{empty}

\section{Introduction\label{introduction}}

The volume of archival data is growing exponentially, driven by factors such as future analytics needs and regulatory compliance requirements ~\cite{borovica2016cheap}. As a result, the world’s archive of stored data continues to expand at a rapid pace. Yet, it is estimated that over 80\% of this archived data will either never be accessed again or only accessed very infrequently. This category of rarely accessed information is referred to as "cold data" ~\cite{storagecold}.

Traditional cold storage solutions, such as hard drives, tape, and optical devices, have limited lifespans — typically between 5 and 20 years, depending on the medium. This limitation means that long-term storage data must be periodically migrated to new media, increasing both costs and complexity over time \cite{balakrishnan2014pelican}.

In light of these limitations, DNA — nature's own information storage system, refined over billions of years — has emerged as a promising medium for long-term data preservation ~\cite{church2012next}. DNA offers unmatched data density and durability. In fact, estimates suggest that all the digital data in the world could theoretically be stored in a space no larger than a shoebox ~\cite{erlich2017dna}. However, despite its immense potential, the prohibitive cost and speed of DNA synthesis remains a significant barrier to its widespread adoption ~\cite{ceze2019molecular}.

To address the high synthesis costs, several approaches have emerged to circumvent the standard base-by-base process. One class of novel synthesis writes data to the DNA with pre-synthesized motifs, which are short subsequences of DNA ~\cite{roquet2021dna}. Doing so simplifies synthesis, reduces its cost and improves downstream decoding. In addition, encoding information over a combination of motifs from a fixed motif library increases the logical information density as well ~\cite{yan2023scaling, preuss2021data}. These are broadly classified as motif-based methods for synthesizing DNA for DNA storage.

To read the data, the motifs must be recovered from the raw signal coming from DNA sequencing devices. Today, high-throughput technologies, such as Nanopore sequencers, are readily accessible, making sequencing more cost-effective \cite{ceze2019molecular, mardis2013next}. Nanopore sequencing involves threading a DNA strand through a Nanopore that has an electrical current running through it \cite{mikheyev2014first}. As the DNA strand passes through, it disrupts the current, generating a series of electrical signal deflections that correspond to the individual bases. Machine learning based algorithms are then used to convert the recorded signal trace into a sequence of DNA bases in a process known as basecalling \cite{wick2019performance}. After basecalling, a motif detection algorithm called Motif Search then finds the motifs in the sequence to decode the data \cite{yan2023scaling}.

This paper explores an innovative approach to shorten the process and proposes a machine learning model --- a Motif Caller --- that directly maps the raw signal to motifs. Doing so means that the machine learning model can work with more features (mapping several bases in one go instead of just one) which improves the ability of the model to detect motifs. We aim to increase the percentage of motifs detected per read, which in turn, leads to a lower effective sequencing coverage. We start by breaking down the two methods of motif-based synthesis (Sec.~\ref{motif_based_synthesis}), the Helixworks approach \cite{yan2023scaling} and the CatalogDNA approach \cite{roquet2021dna}. Following this, we introduce our motif-inferring methods, the Motif Search pipelines (Sec.~\ref{baseline}) and consequently our proposed approach of the Motif Caller (Sec.~\ref{caller}). We then introduce the datasets that we use to compare the methods (Sec.~\ref{dataset}) and evaluate the methods on the datasets (Sec.~\ref{results}). We conclude with a brief discussion of the results, potential further improvements of the Motif Caller (Sec.~\ref{discussion}) and its applications outside of DNA storage. By addressing the key challenge of sequencing efficiency, our research brings us one step closer to realizing the immense potential of DNA as a revolutionary data storage medium.

\section{Motif-Based Synthesis\label{motif_based_synthesis}} 
The practical application of DNA for data storage has historically been constrained by the high cost and low throughput of conventional phosphoramidite chemistry, which synthesizes DNA strands one nucleotide at a time. To overcome this, motif-based synthesis has emerged as a new paradigm where information is encoded by assembling pre-synthesized DNA subsequences, or "motifs" \cite{oa2020prototype, yan2023scaling, roquet2021dna}. The CMOSS system has been developed to conduct motif-based synthesis. While it uses phosphoramidite chemistry, its innovative columnar data layout enables an integrated consensus and decoding pipeline that prevents the propagation of indel errors during seqeuncing - reportedly achieving full data recovery with as little as 4x sequencing coverage \cite{marinelli2024cmoss}

Three distinct architectural philosophies have been reported for conducting motif-based storage, by Catalog, BISHENG-1, and Helixworks, which differ in their assembly mechanics, encoding strategies, and optimization goals. Two of these strategies, from Catalog and BISHENG-1, prioritize massive parallelism and cost reduction through high-throughput assembly. Though they use different encoding schemes, a Cartesian product of motifs for Catalog \cite{roquet2021dna} versus a "movable type" block assembly for BISHENG-1 \cite{wang2024cost}, both are implemented using custom-developed, automated inkjet printers that dispense and ligate prefabricated, double-stranded DNA (dsDNA) fragments with sticky overhangs. The economic potential of this approach is significant; as demonstrated by the BISHENG-1 system, a single synthesis of a motif can support up to 10,000 assembly reactions. This reusability, combined with the potential to scale down droplet volumes from microliters ($\mu$L) to nanoliters (nL), indicates that write costs for such systems can be reduced from \textdollar 122 per megabyte to potentially as low as \textdollar 0.12 per megabyte \cite{wang2024cost}. However, a significant practical limitation of these high-throughput systems is the capital cost associated with both the initial synthesis of the library and the expensive, dedicated inkjet nozzles required for each unique motif.

\begin{figure}[t]
\centering
\includegraphics[width=\textwidth]{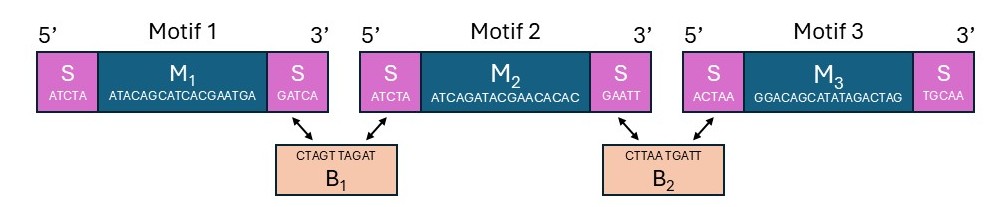}
\caption{Illustration of the typical block-like oligo structure in the Helixworks DNA storage pipeline. Using the bridge ($B_x$), the spacers ($S$) of two motifs are concatenated using their complementary sequences. Motifs are added to the pool one-by-one, and once all the motifs are pooled, the complete oligo is generated in a one-step assembly process.}
\label{fig:dataset}
\end{figure}

The strategy used by Helixworks prioritizes maximizing information density per payload slot \cite{oa2020prototype, yan2023scaling, sokolovskii2024coding}. This system utilizes a different assembly methodology, performing ligations of single-stranded DNA (ssDNA) on a more conventional OpenTrons liquid handling platform, where a "bridge" oligo facilitates the joining of adjacent motifs. This is shown in fig. ~\ref{fig:dataset}, where the procedure of this ligation using a bridge oligo is demonstrated. The resulting oligo architecture is payload dominant, comprising a small number of identifier slots (three address slots) followed by eight payload slots to form an oligo of approximately 625~bp. Information is encoded using a "composite motif" technique where, for each of the eight payload slots, a single combinatorial symbol is represented by a mixture from a subset of motifs (4 chosen from a library of 8). This yields a logical density of $\log_{2}\binom{8}{4} \approx 6.1$ bits per synthesis cycle/slot.

Despite their differences, all the systems are dependent on the accurate identification of their constituent motifs from noisy sequencing data. An error in identifying a single motif can corrupt a data symbol, invalidate an address, or select an incorrect block, leading to data loss. This common point of failure highlights the universal need for a robust motif identification tool that is agnostic to the overarching oligo architecture. The conventional two-step decoding pipeline, basecalling raw signals to nucleotides and then searching for motif sequences, is suboptimal as it discards valuable signal-level information. A tool like Motif Caller, which directly interprets raw signals to identify entire motifs, is therefore broadly applicable and essential for improving the fidelity of any motif-based storage system.

The data generated by the the Helixworks' multi-slot, composite motif architecture was used to develop and test the Motif Caller. The system's reliance on sequential combinatorial encoding presents a particularly demanding and relevant use case for direct motif detection. The high logical density achieved through composite motifs amplifies the negative impact of identification errors, and the multi-slot structure requires precise sequential decoding, making it an ideal environment to validate the performance gains of bypassing intermediate basecalling.

\begin{figure*}[!b]
\centering
\includegraphics[width=\textwidth]{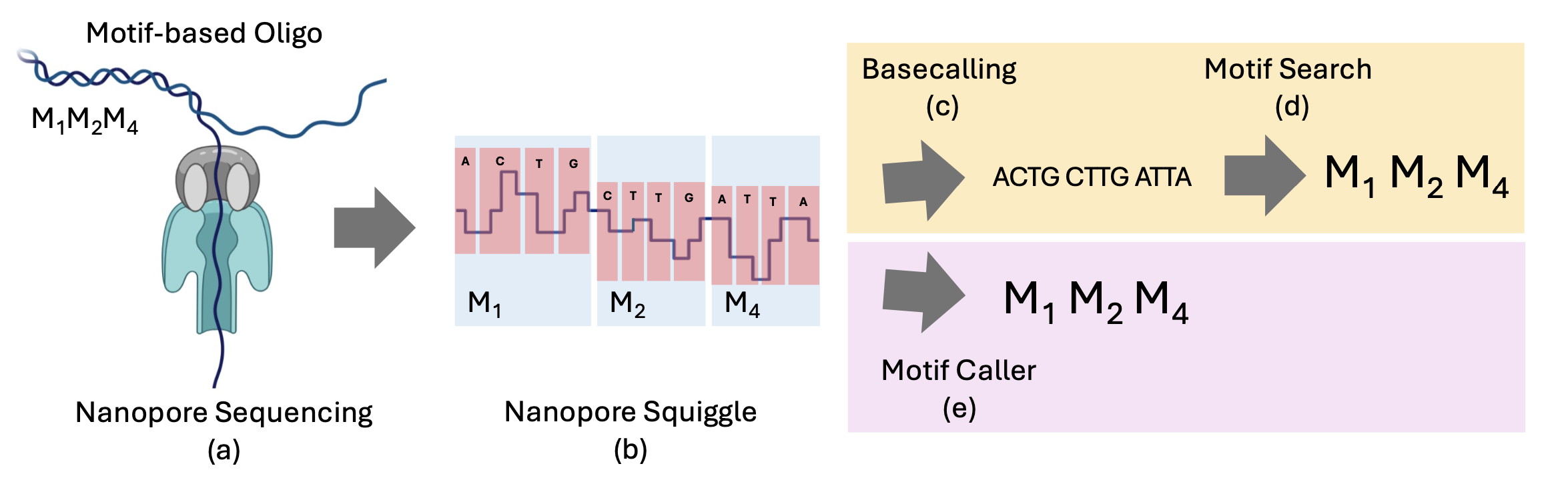}
\caption{Figure illustrating the difference in operation between the Motif Search and Motif Caller pipelines for sequence reconstruction. A motif-based oligo containing the symbols $M_1, M_2, M_4$ passes through the nanopore (a) and produces a trace of the bases in the squiggle (b). The signal is used to predict a base level sequence using basecalling (c). From this base level sequence, Motif Search (d) is used to obtain the final motif sequence prediction. The Motif caller (e) shorts the two-step reconstruction by predicting the motif sequence directly from the squiggle.}
\label{fig:pipeline}
\end{figure*}

\section{Basecalling \& Motif Search - The Baseline} \label{baseline}

Fig.~\ref{fig:pipeline} (a-d) illustrates the baseline, a motif-based DNA storage sequencing pipeline. A motif-based oligo passes through the nanopore (Fig.~\ref{fig:pipeline}a) to produce an electrical current readout, called the squiggle (Fig.~\ref{fig:pipeline}b). This squiggle is the deflection of the current as the bases of the strand pass through the nanopore. The remainder of the pipeline aims to reconstruct the motif sequence from the squiggle in two steps.

Basecalling (Fig.~\ref{fig:pipeline}c) reconstructs the sequence of bases from the squiggle created by the oligo passing through the nanopore. Basecallers learn the representation of individual bases within a squiggle, based on which they make inferences to reconstruct the base-level sequence.

\subsection{Zero-error (ZE) Search}
Once a base-level prediction is made from the squiggle, the original stored motif sequence can be reconstructed. (Fig.~\ref{fig:pipeline}d). The challenge is to infer the motifs from a noisy base-level prediction, which is especially prone to deletion errors. Zero-error motif search \cite{yan2023scaling} looks for exact matches of motifs, which means that the entire sequence of bases in a motif must be basecalled correctly in order for a motif to be detected. The performance of Zero-error search is inherently limited by the ability of the basecaller to predict a sequence of bases correctly. This dependence leads to limited performance, and a large number of deletions in the prediction.

\subsection{Approximate-matching (AM) Search}

Approximate-matching motif search \cite{Yiqing_2022} overcomes this limitation by making the best guess of the motifs at a certain cycle position. First, during segmentation, the algorithm locates candidate spacer positions by matching short k-mers from the read to a pre-built index of the spacer sequences. To handle sequencing errors, it filters out weak candidates, merges nearby candidates, and corrects spacer positions using randomized embedding and Hamming distance comparisons, producing accurate spacer locations. It then identifies chains of spacers spaced according to the expected motif-spacer structure, allowing for small variations due to insertions or deletions.

Next, in the mapping step, AM search extracts segments between the spacer positions and aligns them to a reference library of the motifs that were used for storing information. This is done using a sequence alignment library (ksw-lib \cite{suzuki2018introducing}), which selects the best-matching motif for each segment. Finally, an overlap check ensures that each nucleotide in the read is assigned to only one motif/oligo: overlapping chains are grouped, and only the chain with the highest mapping score in each group is retained. This process enables accurate inference of the motif sequence despite sequencing errors and read variability.

\section{Motif Caller} \label{caller}

The Motif Caller pipeline, illustrated in Fig.~\ref{fig:pipeline} (a, b \& e), attempts to shorten the two step process of basecalling followed by Motif Search by predicting a motif-level sequence directly from the squiggle. With perfect data, we would expect a motif level accuracy similar to basecallers, or even superior - owing to a larger number of features per motif as compared to a base.

\begin{wrapfigure}{r}{0.24\textwidth}
  \begin{center}
    \includegraphics[width=0.22\textwidth]{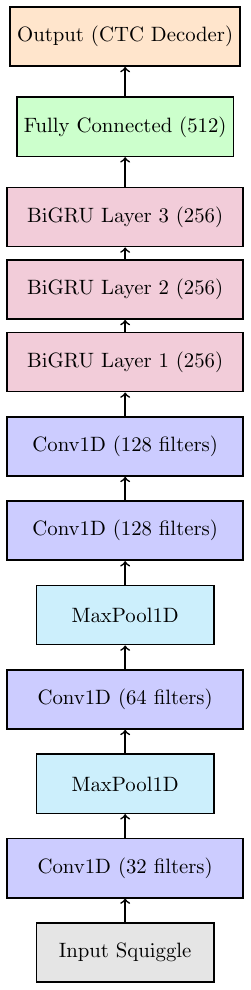}
  \end{center}
  \caption{Model architecture used for the Motif Caller. The architecture is similar to models used for basecalling such as Guppy and Dorado.}

  \label{fig:model_arc}
\end{wrapfigure}

However, generating perfect labels in a real world scenario is challenging, especially in the combinatorial pipeline, since the synthesis process leads to uncertainty about the motifs within a particular block (as we elaborate further in Sec.~\ref{dataset}). We expect the traditional basecalling architecture to generalize well to this use-case. The difference with the standard basecalling models is the larger size of the motif (25 nt) and a larger motif library (8 motifs as compared to 4 bases). We use a similar training strategy to the standard basecalling models such as Dorado, Chiron and Guppy \cite{wick2019performance}. The model uses connectionist temporal classification (CTC) loss in order to operate directly with unsegmented raw data \cite{teng2018chiron}. The model architecture is shown in Fig~\ref{fig:model_arc}. The convolutional layers extract important features from raw data and feed it to the sequential layers of the model, that are bidirectional gated recurrent units \cite{schuster1997bidirectional}. These approximate the function that conduct the prediction for each squiggle (current signal), whilst being able to deal with the temporal dependencies of the labels. The CTC loss helps the model to learn the alignment between the squiggles and the motifs while it trains, without any explicit event segmentation.

\subsection{Training Procedure}

We aim to predict the motif sequence directly from raw squiggle data without any segmentation steps. Let $M$ denote the original set of motifs used to store information, comprising 8 distinct motifs. Each motif $m \in M$ is a random sequence of bases of length $l$. In our datasets, all motifs have the same length of $l=25$ bases, without loss of generality. The task is to map the squiggle sequence input to their corresponding motif labels. Let the raw signal input be,  $s = [s_1, s_2, \ldots, s_T]$, which needs to be mapped to their respective motif labels $y = [y_1, y_2, \ldots, y_k]$. Training samples are drawn from the dataset \( X = \{ (s^{(1)},  y^{(1)}), (s^{(2)},  y^{(2)}), \ldots \} \). By using the model architecture and loss function described below, the input time series \( s \) can be directly translated to the sequence \( y \) without any segmentation steps.

For each training sample, the convolutional layers extract local features from the raw signal, generating a sequence of feature maps. These are progressively downsampled by a factor of 64 through a combination of two convolutional layers with stride 2 and two max-pooling layers with stride 4. The resulting feature sequence is then passed to a stack of Gated Recurrent Units (GRUs), which capture the temporal dependencies present in the signal. The output of the GRUs is fed into a fully connected layer that maps the temporal representations to a set of label classes. The model outputs a series of probability distributions, which are the probabilities of observing each token over a number of timesteps. The number of timesteps are about 3 times the number of targets in the label. CTC introduces an additional "blank" token, denoted as \( \phi \), to the set of possible labels. This token allows the model to learn the transitions between the tokens.
From the model output, a sequence of labels \( \pi = [\pi_1, \pi_2, \ldots, \pi_T] \) can be produced, by selecting a token at each window. Each $\pi_i$ is one of the motifs in $M$ or $\phi$. This sequence of labels is called an alignment. Each alignment \( \pi \) must be mapped to the target sequence \( y \).  This is done by first removing all the repeated characters, and then dropping the blanks. For example, the alignment \( [A, \phi, A, A, \phi, G, G] \) maps to the sequence \( [A, A, G] \).  The probability of an alignment \( \pi \) given the input sequence \( s \) is the product of the probabilities of the labels at each window $t$: \[P(\pi | s) = \prod_{t=1}^{T} P(\pi_t | s)\]. The total probability of the target sequence \( y \) given the input sequence \( s \) is the sum of the probabilities of all possible alignments that map to \( y \): \[P(y | s) = \sum_{\pi \in \mathcal{B}(y)} P(\pi | s)\] where \( \mathcal{B}(y) \) is the set of all alignments that collapse to \( y \). The CTC loss is the negative log probability of the correct label sequence: \[\mathcal{L}_{CTC} = -\log P(y | s)\]

This encourages the model to maximize the probability of the correct target sequence after considering all possible alignments. The CTC criterion computes these alignments using a dynamic programming algorithm \cite{graves2006connectionist}, that allows dealing with large sequences without causing an explosion in computational complexity.

\subsection{Decoding}

During inference, the final sequence is constructed using either a greedy decoder or a beam-search decoder, following standard techniques as described in \cite{teng2018chiron}. The greedy decoder selects the token of maximum probability at each timestep and collapses the alignment by removing all repeated characters and dropping the blanks. The beam search decoder (with beam width $W$), maintains a list of the $W$ most probable sequences at each timestep. For the following timestep, it constructs the probability of possible extensions of the sequences by collapsing and summing up over repeated bases or repeated blanks that are terminated by non-blanks.

In the specific case of this dataset, each payload is separated by unique spacer motifs, which eliminates the need to model transitions between repeated characters. For this reason, the difference between greedy and beam search decoding becomes negligible. Therefore, greedy decoding is used during inference, while accounting for quality thresholds\cite{ewing1998base}, which are calculated as:
\begin{equation}
    Q = -10 \times log_{10}(1 - P_{C})
\end{equation}

Where $P_C$ denotes the model confidence of each predicted token. Quality thresholding is applied both at the read level (via average quality score) and the token level (by filtering out low-confidence predictions). In the Motif Search pipelines, basecalled reads with quality scores below 10 are excluded. Similarly, in the Motif Caller pipeline, reads with average quality scores below 11 are filtered out to match the primary dataset’s criteria. Additionally, within the Motif Caller, individual predicted motifs with confidence scores below 0.85 are treated as blanks and discarded.

\section{Dataset} \label{dataset}

The empirical dataset used to train the Motif Caller model was obtained from an experimental run conducted by Helixworks (Sec.~\ref{motif_based_synthesis}) who have developed a combinatorial approach to motif-based DNA storage \cite{yan2023scaling, sokolovskii2024coding}. 

\begin{wrapfigure}{r}{0.45\textwidth}
  \begin{center}
  \centering
  \includegraphics[width=0.42\textwidth, height=0.21\textwidth]{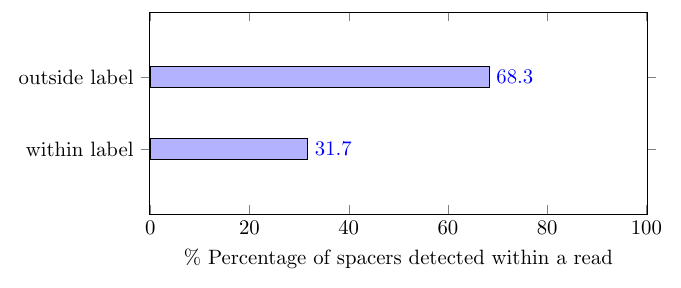}
  \end{center}
  \caption{The Motif Caller successfully learns to identify spacer motifs that are not explicitly present in its training labels, indicating its ability to generalize beyond the provided supervision even during training.}
  \label{fig:generalise_motifs}
\end{wrapfigure}

The primary dataset comprises $\approx 4000$ blocks of information, with an average of $\approx 200$ nanopore reads per cycle. Each block consists of one address motif and eight payload (information-carrying) motifs which are separated by position-specific spacer sequences. All motifs are 25 bases in length.

Due to the combinatorial nature of the synthesis pipeline, there is an inherent uncertainty in determining which specific motifs are within a particular information block. This creates a clear distinction between the original information that is stored (i.e., the \textit{pre-synthesis ground truth}), and the actual motifs present in the resulting DNA strands. As a result, generating accurate labels for training the Motif Caller becomes non-trivial, since the true motif sequence within each read is unknown.

To construct training labels, we combine predictions from the Approximate-matching (AM) Motif search pipeline (Sec.~\ref{baseline}) with the pre-synthesis ground truth. The spacer motifs are used to segregate the payload motifs into their cycle positions. The reads that have the largest proportion of detected motifs matching the ground truth are selected. Specifically, the top $30 \%$ of the reads from the AM search pipeline are retained for training. Since the labels are generated in this fashion, the performance of the Motif Caller is influenced by the best performance of the Motif Search pipeline, where it detects $\approx$ $90 \%$ of the motifs within a read. Despite this influence, as we show in Fig. \ref{fig:generalise_motifs}, the Motif Caller learns the spacer motifs that are not present within the labels, showing that the model is able to generalize outside of its training labels.

To further evaluate the model’s generalization capabilities, we test it in two distinct scenarios. First, from the EIC04 sequencing run, four datasets are created with varying DNA concentrations per run, resulting in different sequencing coverage levels per information block. Dataset T1, with the greatest sequencing coverage, is used to generate labels and train the Motif Caller. The model is then tested on datasets T2, T3, and T4, which exhibit progressively lower sequencing coverage, to assess decoding accuracy under limited read conditions.
Second, we evaluate the model on data from three independent sequencing runs — 01-04, 01-13, and 01-14. These runs follow the same combinatorial design and constraints; however, differences in sequencing conditions may affect the average quality of the resulting reads. For this evaluation, a subset comprising approximately 10\% of the total information blocks is sampled from each run to compare motif inference performance across methods.

\section{Results} \label{results}
To evaluate the Motif Caller pipeline,
the performance of the Motif Caller is compared with the Motif Search methods on the dataset on which the model was trained (Sec.~\ref{dataset}). Following this, the performance of the Motif Caller on data produced by sequencing runs from separate experiments is evaluated. During basecalling, all predictions with quality scores below 10 are filtered out. To maintain consistency with this filtering level, the Motif Caller excludes all predictions with quality scores below 11. The proportions of reads retained after filtering are presented in Table~\ref{table:metrics_summar}.

\begin{figure*}[!b]
\centering
\begin{minipage}{.42\textwidth}
  \centering
  \subcaption[.0.02\textwidth]{\textcolor{white}{This text is invisible on a white background.}}
  \includegraphics[width=.85\linewidth]{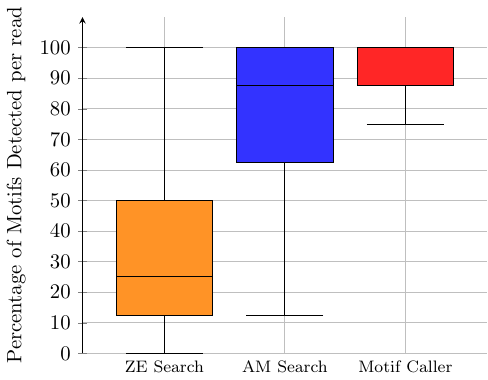}
  \label{fig:box_motifs_detected_empirical}
\end{minipage}%
\begin{minipage}{.42\textwidth}
  \centering
  \subcaption[.0.02\textwidth]{\textcolor{white}{This text is invisible on a white background. Mayabe some more}}
  \includegraphics[width=.78\linewidth]{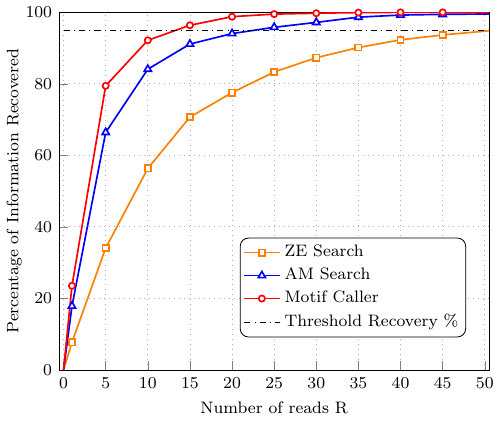}
   
  \label{fig:inf_recovered_empirical}
\end{minipage}
\caption{Results on the primary dataset (EIC04). (a) Comparison of the percentage of motifs identified per read by the Motif Caller and the Motif Search pipelines on unseen information blocks. The Motif Caller consistently detects a higher proportion of motifs per read. (b) Information recovery as a function of the number of reads. The threshold for complete recovery is marked at 95\%, illustrating the lower coverage requirement of the Motif Caller compared to baseline methods.}
\end{figure*}

\subsection{Motif accuracies per read}

The motifs detected per read is the sum of the number of motifs that are detected per payload in a single read, divided by the total number of motifs in an information block (8 such motifs). The spacer motifs are used to segregate the motifs into their respective payloads, and compare these payloads to the pre-synthesis ground truth. Due to the combinatorial nature of the ground truth, there is an uncertainty about the payload motifs that are present within a particular read. We compare the motifs detected per read to the pre-synthesis ground truth.
In Fig.~\ref{fig:box_motifs_detected_empirical}, the percentage of motifs detected between the Motif Caller (blue) and the Motif Search methods - Approximate Matching (blue) and Zero-error (orange) is compared over the reads for all the unseen blocks of information of the empirical dataset. It can be seen that the Motif Caller consistently detects a higher percentage of motifs as compared to the Motif Search methods.

Since ZE search is dependent on exact matches of a whole motif at the base-level, it requires 25 bases to be identified in a row without any errors. Due to the error prone nature of basecalling, this is  unlikely. In contrast, AM search looks for the best approximate prediction from the base-level inference, and thus performs significantly better than ZE search. Nevertheless, the Motif Caller consistently outperforms both methods in terms of motif detection.
In Table ~\ref{table:EIC04_results}, the accuracies of the methods are presented for forward and reverse-oriented reads. All methods perform significantly better on forward reads as compared to reverse reads. Both AM search and the Motif Caller have a higher error rate as compared to ZE search, highlighting a trade-off between detection sensitivity and precision.

\begin{table}[t]
    \centering
    \begin{tabular}{c| c c c |c}
            \hline
            Method & Orientation & Motifs detected (\%)  & Motif Error (\%) & Coverage (reads)\\
            \hline
            & Forward   & 32.6 &   1.0\\
            ZE Search & Reverse   & 26.5 &   1.0 & 51\\
             &Combined & 30.9 &1.0 & \\
            \hline
            & Forward   & 80.9 &   2.8 & \\
            AM Search & Reverse   & 63.8 &   3.3 & 23\\
             & Combined & 75.8 & 3.0 & \\
            \hline
            & Forward   & 94.9 &   2.8 &\\
            Motif Caller & Reverse   & 77.1 &   2.5 & 14\\
            & Combined  & 91.1 & 2.6 & \\
            \hline
        \end{tabular}
        \caption{Performance metrics for different motif detection methods (ZE search, AM search, and Motif Caller) on the EIC04 sequencing run, which served as the training dataset for the Motif Caller. Metrics include the average percentage of motifs detected per read, error rates, and the effective sequencing coverage—defined as the number of reads required to achieve 95\% information recovery.}
        \label{table:EIC04_results}
    
\end{table}

\begin{comment}
\begin{figure}[!b]
    \centering
    \includegraphics[width=0.62 \linewidth]{Figures/results/gen_results_bar_plot.pdf}
    \caption{Percentage of motifs detected by different methods (AM search, ZE search, and Motif Caller) across multiple sequencing runs. The Motif Caller consistently detects more motifs than the baseline methods, demonstrating its strong generalization. EIC04 was the primary dataset used for generating labels for training the Motif Caller.}
    \label{fig:enter-label}
\end{figure}
\end{comment}

\subsection{Information recovered per block with increasing reads}

To recover all the information that is stored in a block, majority votes are taken over repeated reads. This is done until all the voted motifs in all the payloads of a block match the pre-synthesis ground truth. Since the last few motifs will take an increasing number of reads to be recovered (due to particularly erroneous blocks), the methods approach a recovery close to 100 \% faster than they converge to the full information. In practice this would be intervened with an appropriate error-correcting scheme \cite{sokolovskii2024coding}.

In Fig.~\ref{fig:inf_recovered_empirical}, we compare this information recovery averaged over all the blocks of information, plotted for the Motif Caller (red with circle markers), the AM Search (blue with triangle markers) and ZE Search (orange with square markers).
We mark a recovery close to 100\% in order to investigate how many reads the methods take to reach an arbitrary threshold before converging to the full information.

 The Motif Caller reaches the threshold recovery at 14 reads, AM search reaches the threshold at 23 reads and ZE search reaches the threshold at 51 reads. Even with a $15\%$ difference in the motifs detected per read, the overall sequencing coverage required can be reduced by 30\%. We present these metrics in Table ~\ref{table:EIC04_results}, where we compare the motifs detected and the effective sequencing coverage between the Motif Caller and Motif Search. By training the Motif Caller on the set of motifs, even with imperfect labels, we can significantly reduce the effective sequencing coverage required over the blocks of information stored.

\subsection{Decoding percentage for diluted reads}

\begin{table}[b]
    \centering
    \begin{tabular}{c c c c c}
            \hline
            Method & Decoding accuracy & Dataset & Concentration (amol/block) & Reads / block  \\
            \hline
            ZE Search & 71.6& &  & \\
             AM Search & 89.9& T2& 1.07& 18.0\\
            Motif Caller & 96.9 & & & \\
            \hline
            ZE Search & 41.6& &  & \\
             AM Search & 64.9& T3& 0.44& 7.1\\
            Motif Caller & 71.8 & & & \\
            \hline
            ZE Search & 17.4& &  & \\
             AM Search & 36.0& T4& 0.11& 2.3\\
            Motif Caller & 38.7 & & & \\
            \hline
        \end{tabular}
        \caption{Performance of the different motif detection methods (ZE search, AM search, and Motif Caller) on the dataset with varying concentrations from the EIC04 sequencing run. The Motif Caller achieves higher decoding accuracy across all datasets.}
        \label{table:EIC04_diluted}
    
\end{table}

For the EIC04 run, three additional sequencing runs were performed using progressively lower concentrations of DNA (Sec. ~\ref{dataset}), resulting in varying numbers of reads per information block. Table~\ref{table:EIC04_diluted} presents the decoding accuracy (\%) of the different motif detection methods—ZE Search, AM Search, and Motif Caller—on the diluted datasets T2, T3, and T4.

As sequencing coverage decreases, all methods show a decline in decoding accuracy; however, the Motif Caller consistently outperforms both ZE Search and AM Search across all datasets. For instance, on dataset T2 (1.07 amol/block, 18.0 reads/block), the Motif Caller achieves 96.9\% accuracy compared to 89.9\% for AM Search and 71.6\% for ZE Search. Even at the lowest coverage in T4 (0.11 amol/block, 2.3 reads/block), the Motif Caller maintains a performance advantage, demonstrating its robustness under low-read conditions.

\subsection{Generalization of performance to alternative sequencing runs}

To evaluate the generalization ability of the model, it was tested on data from three independent sequencing runs: 01-04, 01-13, and 01-14. Reads with a basecalling quality score below 10 were filtered out in the Motif Search pipelines, while for consistency with the primary dataset, reads with quality scores below 11 were filtered out in the Motif Caller pipeline.

\begin{wrapfigure}{r}{0.44\textwidth}
  \begin{center}
    \includegraphics[width=0.43\textwidth]{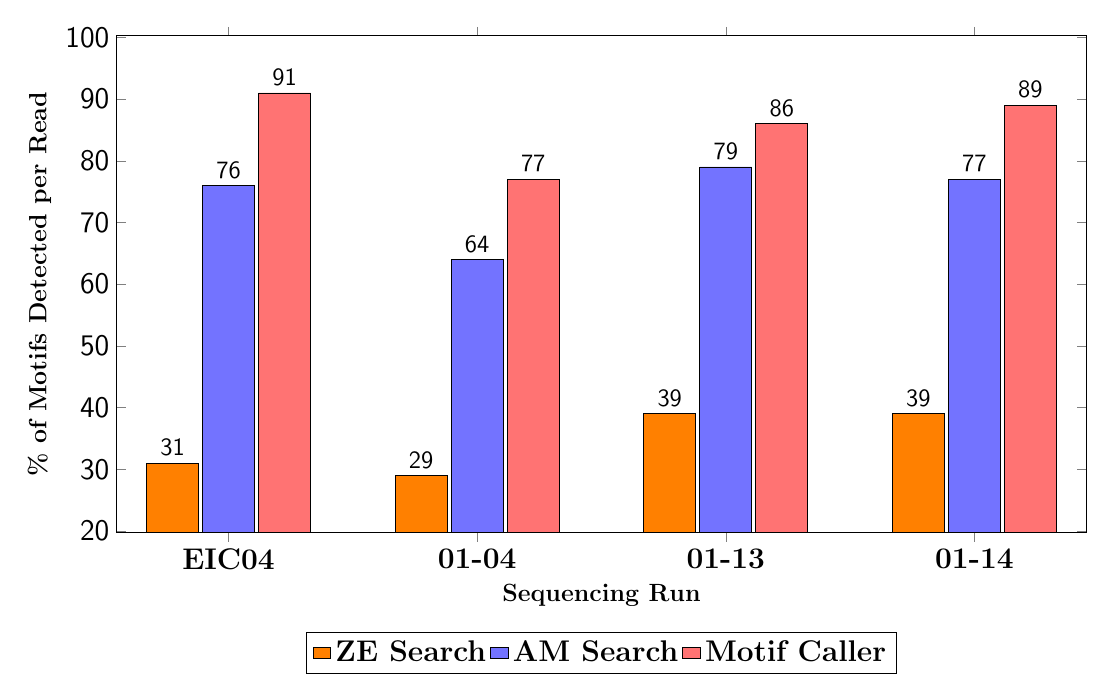}
  \end{center}
  \caption{Percentage of motifs detected by the different motif detection methods (AM search, ZE search, and Motif Caller) across multiple sequencing runs. The Motif Caller consistently detects more motifs than the baseline methods, demonstrating its strong generalization. EIC04 was the primary dataset used for generating labels for training the Motif Caller.}

  \label{fig:gen_results}
\end{wrapfigure}
 
Figure~\ref{fig:gen_results} presents the number of motifs detected across these runs using  the different methods. The Motif Caller (red) consistently outperforms both the AM search (blue) and ZE search (orange) across all sequencing runs. Notably, all methods exhibit reduced performance on the 01-04 run relative to the others. Table ~\ref{table:metrics_summar} summarizes the number of motifs detected, error rates, the effective sequencing coverage required, and the percentage of reads retained after applying the filtering criteria. Consistent with the results from the primary dataset, ZE search consistently achieves a significantly lower error rate compared to both AM search and the Motif Caller, which exhibit comparable error rates.

 The Motif Caller achieves threshold information recovery using a significantly smaller fraction of the reads. For the 01-13 and 01-14 sequencing runs, a significantly larger proportion of reads meet the Motif Caller’s filtering criteria compared to the Motif Search pipelines, thereby further reducing the effective sequencing coverage required. In contrast, only 20\% of reads from the 01-04 run pass the Motif Caller’s quality threshold, which leads to an increased effective sequencing coverage requirement for that dataset.

\begin{table}[t]
    \centering
    \begin{tabular}{c c c c c c c}
            \hline
            Method & Sequencing run 
            & Motifs detected (\%)  & Motif Error (\%) & Coverage (reads) & \% of total reads\\
            \hline
            & EIC04 & 30.9 & 1.0
            & 51   & 59.8\\
            & 01-04 & 29.0 & 1.2
            & 46   & 61.3\\
            & 01-13 & 38.8 & 1.9
            & 37   & 51.9\\
            ZE Search& 01-14 & 39.0 & 1.6
            & 36   & 67.6\\
            \hline
            & EIC04 & 75.8 & 3.0
            & 23   & 45.6\\
            & 01-04 & 63.8 & 2.3
            & 35   & 48.6\\
            & 01-13 & 78.9 & 3.5
            & 28   & 41.6\\
            AM Search& 01-14 & 77.4 & 4.5
            & 31   & 54.8\\
            \hline
            & EIC04 & 91.1 & 2.6
            & 14   & 71.1\\
            & 01-04 & 77.0 & 3.6
            & 26   & 23.4\\
            & 01-13 & 86.0 & 3.6
            & 17   & 68.3\\
            Motif Caller& 01-14 & 89.1 & 4.0
            & 18   & 72.5\\
            \hline
        \end{tabular}
        \caption{Summary of the key performance metrics for the different motif detection methods (ZE search, AM search and Motif Caller) across multiple sequencing runs. Reported metrics include the average percentage of motifs detected per read, error rates, the effective sequencing coverage required for threshold data recovery (95\%), and the proportion of the reads retained after quality filtering. EIC04 refers to the original sequencing run used to generate labels for the Motif Caller. The Motif Caller achieves higher motif detection rates and lower coverage requirements across more runs, with the exception of 01-04, where fewer reads meet the model's confidence threshold.}
        \label{table:metrics_summar}
    
\end{table}

\subsection{Finetuning and Quality Proportions}

The Motif Caller demonstrates consistent performance on sequencing runs other than the one it was trained on, with the notable exception of the 01-04 run. For this particular dataset, both the motif recovery performance at the quality threshold and the proportion of reads that meet this threshold are substantially lower.
To investigate whether model performance could be improved through adaptation, the Motif Caller was fine-tuned for 15 epochs on a subset of 30,000 reads from the 01-04 sequencing run, spanning various blocks of information. As shown in Table~\ref{table:finetuned}, the fine-tuned model exhibits a marked improvement in performance at the quality threshold. However, the percentage of reads that fall within this threshold does not change.
These results suggest that when only a small fraction of reads meet the model’s confidence criteria, retraining the model on data from the specific sequencing run may be necessary to achieve comparable coverage to the original training conditions. Fine-tuning effectively enhances performance for high-confidence reads, without shifting the proportion of reads at the different quality thresholds. In these cases, the Motif Caller may still be used in conjunction with baseline methods such as AM search to reduce the effective coverage requirement, leveraging the model’s confidence scores to guide this integration.

\begin{table}[t]
    \centering
    \begin{tabular}{c c c c c}
            \hline
            Model & Quality Threshold & \% of total reads & Motifs detected (\%)  & Motif Error (\%)\\
            \hline
            & Q0    & 100.0 &   49.8 & 6.0 \\
            & Q10   & 48.4&   70.8 & 5.3 \\
            & Q15   & 19.9&   87.1 & 3.4 \\
            Motif Caller & Q20   & 11.2&   93.6 & 2.6 \\
            \hline
            & Q0    & 100.0 &   53.0 & 3.5\\
            & Q10   & 45.3 &   76.9 & 3.1 \\
            & Q15   & 18.0 &   90.3 & 2.5 \\
            Finetuned Caller & Q20   & 8.5 &   92.3 & 2.8 \\
            \hline
        \end{tabular}
        \caption{Comparison of quality threshold distributions and corresponding accuracies for the 01-04 sequencing run, evaluated between the original Motif Caller and a version finetuned on data from the same run. Finetuning improves motif detection and reduces error rates at given quality thresholds; however, it does not increase the proportion of reads that meet the confidence threshold criteria.}
        \label{table:finetuned}
    
\end{table}

\section{Discussion} \label{discussion}
In this paper, we introduced \emph{Motif Caller}, an approach for sequence reconstruction for motif-based DNA storage directly from the raw current signals (squiggles). Even when trained on imperfect labels generated by the Approximate-matching motif search pipeline, the Motif Caller is able to detect more motifs per read than the current methods. As a result, it significantly reduces the amount of sequencing coverage needed to recover each block of stored information. This was demonstrated by consistently recovering more information than the other methods across all coverage levels.

The Motif Caller also demonstrates strong generalization across sequencing runs, performing comparably on two alternate datasets despite being trained on a single run. By using model confidence scores as quality thresholds, it is possible to assess post hoc whether the model generalizes effectively to a given sequencing run. For example, in the 01-04 sequencing run, fewer reads passed the quality threshold, suggesting lower compatibility or higher noise in that dataset. These confidence scores can help decide whether the model needs retraining or if the sequencing run should be repeated.

While the current results are promising, there remain clear avenues for further improvement. First, conducting a synthesis run with known motif payloads would allow the generation of higher-quality ground truth labels, thereby improving the training signal and overall model accuracy. Second, training on a more diverse set of sequencing runs could enhance the model’s robustness and generalizability. Finally, incorporating a structured penalty term during training that reflects prior knowledge of valid payload motifs may further reduce prediction error, even when the exact motif is unknown.

Beyond DNA storage, the Motif Caller holds potential for applications in biological sequence analysis. In particular, the problem of motif detection in natural genomic data shares structural similarities, as it involves identifying patterns from noisy, base-level signals \cite{das2007survey}. As demonstrated in this study, direct inference of motifs from squiggles can outperform traditional basecalling approaches when supported by high-quality training data, which is increasingly available in biological settings due to advances in sequencing technology .

The Motif Caller represents a significant advancement in the sequence reconstruction pipeline for motif-based DNA storage. Together with ongoing progress in error-correcting codes and DNA synthesis, this work contributes to making DNA-based storage systems more practical, scalable, and efficient.

\subsection*{Competing interests}
The authors declare no competing interests.

\subsection*{Data Availability}
The data supporting the findings of this study are available in the Open Science Framework repository at \href{https://osf.io/pcdtj/}{https://osf.io/pcdtj/}

\subsection*{Code Availability}
The code used to develop, train, and evaluate Motif Caller is publicly available at:
\href{https://github.com/Parvfect/Motif-Calling}{https://github.com/Parvfect/Motif-Calling}.

\bibliography{article}

\begin{thebibliography}{10}
\urlstyle{rm}
\expandafter\ifx\csname url\endcsname\relax
  \def\url#1{\texttt{#1}}\fi
\expandafter\ifx\csname urlprefix\endcsname\relax\def\urlprefix{URL }\fi
\expandafter\ifx\csname doiprefix\endcsname\relax\def\doiprefix{DOI: }\fi
\providecommand{\bibinfo}[2]{#2}
\providecommand{\eprint}[2][]{\url{#2}}

\bibitem{borovica2016cheap}
\bibinfo{author}{Borovica-Gaji{\'c}, R.}, \bibinfo{author}{Appuswamy, R.} \& \bibinfo{author}{Ailamaki, A.}
\newblock \bibinfo{journal}{\bibinfo{title}{Cheap data analytics using cold storage devices}}.
\newblock {\emph{\JournalTitle{Proceedings of the VLDB Endowment}}} \textbf{\bibinfo{volume}{9}}, \bibinfo{pages}{1029--1040} (\bibinfo{year}{2016}).

\bibitem{storagecold}
\bibinfo{author}{Storage, C.-b.~C.}
\newblock \bibinfo{title}{Cold storage in the cloud: Trends, challenges, and solutions}.
\newblock \bibinfo{howpublished}{\url{https://www.intel.com/content/dam/www/public/us/en/documents/white-papers/cold-storage-atom-xeon-paper.pdf}}.
\newblock \bibinfo{note}{Accessed: 2024-12-20}.

\bibitem{balakrishnan2014pelican}
\bibinfo{author}{Balakrishnan, S.} \emph{et~al.}
\newblock \bibinfo{title}{Pelican: A building block for exascale cold data storage}.
\newblock In \emph{\bibinfo{booktitle}{11th USENIX Symposium on Operating Systems Design and Implementation (OSDI 14)}}, \bibinfo{pages}{351--365} (\bibinfo{year}{2014}).

\bibitem{church2012next}
\bibinfo{author}{Church, G.~M.}, \bibinfo{author}{Gao, Y.} \& \bibinfo{author}{Kosuri, S.}
\newblock \bibinfo{journal}{\bibinfo{title}{{Next-generation digital information storage in DNA}}}.
\newblock {\emph{\JournalTitle{Science}}} \textbf{\bibinfo{volume}{337}}, \bibinfo{pages}{1628--1628} (\bibinfo{year}{2012}).

\bibitem{erlich2017dna}
\bibinfo{author}{Erlich, Y.} \& \bibinfo{author}{Zielinski, D.}
\newblock \bibinfo{journal}{\bibinfo{title}{{DNA Fountain enables a robust and efficient storage architecture}}}.
\newblock {\emph{\JournalTitle{Science}}} \textbf{\bibinfo{volume}{355}}, \bibinfo{pages}{950--954} (\bibinfo{year}{2017}).

\bibitem{ceze2019molecular}
\bibinfo{author}{Ceze, L.}, \bibinfo{author}{Nivala, J.} \& \bibinfo{author}{Strauss, K.}
\newblock \bibinfo{journal}{\bibinfo{title}{{Molecular digital data storage using DNA}}}.
\newblock {\emph{\JournalTitle{Nature Reviews Genetics}}} \textbf{\bibinfo{volume}{20}}, \bibinfo{pages}{456--466} (\bibinfo{year}{2019}).

\bibitem{roquet2021dna}
\bibinfo{author}{Roquet, N.} \emph{et~al.}
\newblock \bibinfo{journal}{\bibinfo{title}{{DNA-based data storage via combinatorial assembly}}}.
\newblock {\emph{\JournalTitle{bioRxiv}}} \bibinfo{pages}{2021--04} (\bibinfo{year}{2021}).

\bibitem{yan2023scaling}
\bibinfo{author}{Yan, Y.}, \bibinfo{author}{Pinnamaneni, N.}, \bibinfo{author}{Chalapati, S.}, \bibinfo{author}{Crosbie, C.} \& \bibinfo{author}{Appuswamy, R.}
\newblock \bibinfo{journal}{\bibinfo{title}{{Scaling logical density of DNA storage with enzymatically-ligated composite motifs}}}.
\newblock {\emph{\JournalTitle{Scientific Reports}}} \textbf{\bibinfo{volume}{13}}, \bibinfo{pages}{15978} (\bibinfo{year}{2023}).

\bibitem{preuss2021data}
\bibinfo{author}{Preuss, I.}, \bibinfo{author}{Rosenberg, M.}, \bibinfo{author}{Yakhini, Z.} \& \bibinfo{author}{Anavy, L.}
\newblock \bibinfo{journal}{\bibinfo{title}{Efficient dna-based data storage using shortmer combinatorial encoding}}.
\newblock {\emph{\JournalTitle{Scientific reports}}} \textbf{\bibinfo{volume}{14}}, \bibinfo{pages}{7731} (\bibinfo{year}{2024}).

\bibitem{mardis2013next}
\bibinfo{author}{Mardis, E.~R.}
\newblock \bibinfo{journal}{\bibinfo{title}{Next-generation sequencing platforms}}.
\newblock {\emph{\JournalTitle{Annual Review of Analytical Chemistry}}} \textbf{\bibinfo{volume}{6}}, \bibinfo{pages}{287--303} (\bibinfo{year}{2013}).

\bibitem{mikheyev2014first}
\bibinfo{author}{Mikheyev, A.~S.} \& \bibinfo{author}{Tin, M.~M.}
\newblock \bibinfo{journal}{\bibinfo{title}{{A first look at the Oxford Nanopore MinION sequencer}}}.
\newblock {\emph{\JournalTitle{Molecular Ecology Resources}}} \textbf{\bibinfo{volume}{14}}, \bibinfo{pages}{1097--1102} (\bibinfo{year}{2014}).

\bibitem{wick2019performance}
\bibinfo{author}{Wick, R.~R.}, \bibinfo{author}{Judd, L.~M.} \& \bibinfo{author}{Holt, K.~E.}
\newblock \bibinfo{journal}{\bibinfo{title}{{Performance of neural network basecalling tools for Oxford Nanopore sequencing}}}.
\newblock {\emph{\JournalTitle{Genome biology}}} \textbf{\bibinfo{volume}{20}}, \bibinfo{pages}{1--10} (\bibinfo{year}{2019}).

\bibitem{oa2020prototype}
\bibinfo{author}{Pinnamaneni, N.}, \bibinfo{author}{Chalapati, S.} \emph{et~al.}
\newblock \bibinfo{title}{Prototype of enzymatic dna synthesis approach}.
\newblock \bibinfo{type}{Tech. Rep.} \bibinfo{number}{Deliverable 4.1}, \bibinfo{institution}{OLIGOARCHIVE Project, Grant agreement ID: 863320}, \bibinfo{address}{European Commission} (\bibinfo{year}{2020}).
\newblock \doiprefix\url{10.3030/863320}.

\bibitem{marinelli2024cmoss}
\bibinfo{author}{Marinelli, E.}, \bibinfo{author}{Yan, Y.}, \bibinfo{author}{Magnone, V.}, \bibinfo{author}{Barbry, P.} \& \bibinfo{author}{Appuswamy, R.}
\newblock \bibinfo{title}{Cmoss: A reliable, motif-based columnar molecular storage system}.
\newblock In \emph{\bibinfo{booktitle}{Proceedings of the 2024 ACM SIGMOD/PODS Conference on Management of Data}}, \bibinfo{pages}{1--26} (\bibinfo{year}{2024}).

\bibitem{wang2024cost}
\bibinfo{author}{Wang, C.}, \bibinfo{author}{Wei, D.}, \bibinfo{author}{Wei, Z.} \emph{et~al.}
\newblock \bibinfo{journal}{\bibinfo{title}{Cost-effective dna storage system with dna movable type}}.
\newblock {\emph{\JournalTitle{Advanced Science}}} \bibinfo{pages}{2411354} (\bibinfo{year}{2024}).

\bibitem{sokolovskii2024coding}
\bibinfo{author}{Sokolovskii, R.}, \bibinfo{author}{Agarwal, P.}, \bibinfo{author}{Croquevielle, L.~A.}, \bibinfo{author}{Zhou, Z.} \& \bibinfo{author}{Heinis, T.}
\newblock \bibinfo{journal}{\bibinfo{title}{Coding over coupon collector channels for combinatorial motif-based dna storage}}.
\newblock {\emph{\JournalTitle{IEEE Transactions on Communications}}}  (\bibinfo{year}{2024}).

\bibitem{Yiqing_2022}
\bibinfo{author}{Yiqing, Y.}
\newblock \bibinfo{title}{Motif search}.
\newblock In \emph{\bibinfo{booktitle}{https://gitlab.eurecom.fr/yan1/motif-search}} (\bibinfo{year}{2022}).

\bibitem{suzuki2018introducing}
\bibinfo{author}{Suzuki, H.} \& \bibinfo{author}{Kasahara, M.}
\newblock \bibinfo{journal}{\bibinfo{title}{Introducing difference recurrence relations for faster semi-global alignment of long sequences}}.
\newblock {\emph{\JournalTitle{BMC bioinformatics}}} \textbf{\bibinfo{volume}{19}}, \bibinfo{pages}{33--47} (\bibinfo{year}{2018}).

\bibitem{teng2018chiron}
\bibinfo{author}{Teng, H.} \emph{et~al.}
\newblock \bibinfo{journal}{\bibinfo{title}{Chiron: translating nanopore raw signal directly into nucleotide sequence using deep learning}}.
\newblock {\emph{\JournalTitle{GigaScience}}} \textbf{\bibinfo{volume}{7}} (\bibinfo{year}{2018}).

\bibitem{schuster1997bidirectional}
\bibinfo{author}{Schuster, M.} \& \bibinfo{author}{Paliwal, K.~K.}
\newblock \bibinfo{journal}{\bibinfo{title}{Bidirectional recurrent neural networks}}.
\newblock {\emph{\JournalTitle{IEEE Transactions on Signal Processing}}} \textbf{\bibinfo{volume}{45}}, \bibinfo{pages}{2673--2681} (\bibinfo{year}{1997}).

\bibitem{graves2006connectionist}
\bibinfo{author}{Graves, A.}, \bibinfo{author}{Fern{\'a}ndez, S.}, \bibinfo{author}{Gomez, F.} \& \bibinfo{author}{Schmidhuber, J.}
\newblock \bibinfo{title}{Connectionist temporal classification: labelling unsegmented sequence data with recurrent neural networks}.
\newblock In \emph{\bibinfo{booktitle}{Proceedings of the 23rd International Conference on Machine learning}}, \bibinfo{pages}{369--376} (\bibinfo{year}{2006}).

\bibitem{ewing1998base}
\bibinfo{author}{Ewing, B.} \& \bibinfo{author}{Green, P.}
\newblock \bibinfo{journal}{\bibinfo{title}{Base-calling of automated sequencer traces using phred. ii. error probabilities}}.
\newblock {\emph{\JournalTitle{Genome research}}} \textbf{\bibinfo{volume}{8}}, \bibinfo{pages}{186--194} (\bibinfo{year}{1998}).

\bibitem{das2007survey}
\bibinfo{author}{Das, M.~K.} \& \bibinfo{author}{Dai, H.-K.}
\newblock \bibinfo{journal}{\bibinfo{title}{{A survey of DNA motif finding algorithms}}}.
\newblock {\emph{\JournalTitle{BMC bioinformatics}}} \textbf{\bibinfo{volume}{8}}, \bibinfo{pages}{1--13} (\bibinfo{year}{2007}).

\end{thebibliography}

\end{document}